\documentclass[prl,twocolumn,showpacs,superscriptaddress,preprintnumbers]{revtex4}
\usepackage{graphicx}
\usepackage{epstopdf}
\usepackage{dcolumn}
\usepackage{amsmath}
\usepackage{color}

\begin{document}

\title{First-principles Calculation of Atomic Forces and Structural Distortions in Strongly Correlated Materials}

\author{I. Leonov}
\affiliation{Theoretical Physics III, Center for Electronic Correlations and Magnetism,
Institute of Physics, University of Augsburg, 86135 Augsburg, Germany}
\author{V. I. Anisimov}
\affiliation{Institute of Metal Physics, S. Kovalevskaya St. 18, 620219 Yekaterinburg
GSP-170, Russia}
\affiliation{Ural Federal University, 620002 Yekaterinburg, Russia}
\author{D. Vollhardt}
\affiliation{Theoretical Physics III, Center for Electronic Correlations and Magnetism,
Institute of Physics, University of Augsburg, 86135 Augsburg, Germany}

\begin{abstract}
We introduce a novel computational approach for the investigation of complex correlated
electron materials which makes it possible to evaluate interatomic forces and thereby
determine atomic displacements and structural transformations induced by electronic
correlations. It combines \textit{ab initio} band structure and dynamical mean-field 
theory and is implemented with the linear-response formalism regarding atomic displacements. 
We apply this new technique to explore structural transitions of prototypical correlated 
systems such as elemental hydrogen, SrVO$_3$, and KCuF$_3$.
\end{abstract}

 \maketitle



A unified description of the electronic structure and the lattice properties of 
correlated materials is one of the most important goals of modern condensed matter 
physics. This is particularly desirable for correlated electron materials where the 
complex interplay between electronic and lattice degrees of freedom leads to rich 
phase diagrams which makes them interesting not only for fundamental research but 
also for technological applications \cite{Rev}. Clearly, there is a great need for 
theoretical approaches which are able to compute the properties of such materials 
from first principles.

State-of-the-art techniques for the calculation of electronic band structures based on 
density  functional theory often provide a good quantitative description of the electronic 
and lattice properties of a variety of transition metals and semiconductors. The lattice 
structure of such materials can be obtained, e.g., on the basis of total-energy calculations.
These investigations are computationally very demanding even for simple materials,
since they require the minimization of the total energy as a function of all atomic displacements.
The computational effort thus increases exponentially, which limits the applicability of this 
line of approach. Such a problem does not occur when the lattice structure is calculated by 
means of the \emph{forces} acting on the atoms. Indeed, only by computing the complete set of 
interatomic forces using the Hellmann-Feynman theorem \cite{HFT} is it possible to calculate 
the lattice structure of complex materials.
However, even this approach does not lead to satisfactory results in the case of 
\emph{correlated} materials such as the metals Ce, Pu, or the Mott-Hubbard insulators 
NiO, MnO. Extensions of the local density approximation (LDA) or the generalized-gradient 
approximation (GGA), such as the LDA+$U$ approach \cite{LA95}, can considerably improve 
results, e.g. for the band gaps and magnetic moments, but only for magnetically ordered 
states.

The LDA+DMFT approach, a combination of  {\it ab initio} band-structure methods with 
the dynamical mean-field theory (DMFT) of correlated electrons \cite{DMFT} has made 
it possible to compute even the electronic structure of complex, correlated materials 
\cite{DMFT_method}, thereby providing important insights into our understanding of 
their properties \cite{DMFT_calc,SrVO3,JK07,MH04,SKA01,XD03,LB08,LP11,PM13}.
In particular, employing this technique one is now able to study materials located 
at the proximity of a Mott-Hubbard metal-insulator transition as encountered in many 
transition-metal oxides. Moreover, LDA+DMFT calculations can determine both the 
electronic structure and lattice properties of correlated materials, irrespective 
of whether they are in a paramagnetic or a magnetically ordered state 
\cite{MH04,SKA01,XD03,LB08,LP11,PM13}.
Although these studies take the lattice into account only on the basis of total-energy 
calculations, they already demonstrate the crucial importance of electronic correlations 
for the structural stability of correlated materials \cite{SKA01,LP11}.
Furthermore, by implementing the LDA+DMFT scheme within the linear-response formalism
\cite{XD03} one can now investigate the influence of electronic correlations on \emph{dynamical} 
properties of the lattice. Indeed, this approach provides a good quantitative description of the 
electronic properties and lattice dynamics of correlated metals and insulators \cite{SKA01,XD03}.
Nevertheless the approximations needed to solve the quantum impurity model (e.g., the Hubbard 
I approximation) prevent this method to be applicable to correlated materials near a Mott 
metal-insulator transition. Therefore even today it remains a great theoretical challenge 
to treat the electronic and lattice properties of correlated materials in a nonperturbative 
and thermodynamically consistent way.


In this Letter we present a new approach for the calculation of interatomic forces 
and structural distortions in strongly correlated materials based on the implementation 
of LDA+DMFT within the linear-response formalism. The calculation of forces opens the 
way to compute atomic displacements and determine equilibrium atomic positions and, 
hence, explain the origin of lattice transformations induced by electronic correlations.
This makes it possible to calculate the equilibrium lattice structure of correlated 
systems even in the vicinity of a Mott metal-insulator transition --- a computation
which was not feasible up to now.

We start from the total energy functional of a correlated system
\cite{MH04,LB08}
\begin{eqnarray}
\label{eq:energy_functional}
E &=& E_\mathrm{LDA}[\rho] + \langle {\hat H_\mathrm{LDA}} \rangle
- \sum_{m,k}\epsilon^\mathrm{LDA}_{m,k}  \notag \\
&+& \frac{1}{2} \sum_{imm',\sigma\sigma'}
U^{\sigma \sigma'}_{mm'} \langle \hat n_{im\sigma} \hat n_{im'\sigma'}\rangle - E_\mathrm{DC}.
\end{eqnarray}
Here $E_\mathrm{LDA}[\rho]$ denotes the total energy obtained by LDA,
${\hat H_\mathrm{LDA}}$ is the effective low-energy Hamiltonian obtained
from the LDA band structure by employing a projection
technique to evaluate the atomic-centered symmetry-constrained Wannier
orbitals \cite{MV97,AK05,TL08}, $\langle {\hat H_\mathrm{LDA}} \rangle$
is evaluated as the thermal average of ${\hat H_\mathrm{LDA}}$, and
$\sum_{m,k}\epsilon^\mathrm{LDA}_{m,k}$ is the sum of the valence-state
eigenvalues. The interaction energy, the 4-th term on the right-hand side
of Eq.~\ref{eq:energy_functional}, is computed from the double occupancy matrix
$\langle \hat n_{im\sigma} \hat n_{im'\sigma'}\rangle$ which is evaluated in DMFT.
The double-counting correction $E_\mathrm{DC}= \frac{1}{2} \sum_{imm',\sigma\sigma'}
U^{\sigma \sigma'}_{mm'} \langle \hat n_{im\sigma} \rangle \langle \hat n_{im'\sigma'} \rangle$
corresponds to the average Coulomb repulsion between the interacting electrons and is
calculated from the self-consistently determined local occupations.

To evaluate the correlation induced atomic displacements, we calculate the
force acting on the atom $s$ from the first-order derivative of the total energy
\begin{eqnarray}
\label{eq:force_functional}
F_s &=& F^s_\mathrm{LDA} - \delta_s \langle {\hat H_\mathrm{LDA}} \rangle
+ \sum_{m,k} \delta_s \epsilon^\mathrm{LDA}_{m,k}  \notag \\
&-& \frac{1}{2} \sum_{imm',\sigma\sigma'}
U^{\sigma \sigma'}_{mm'} \delta_s \langle \hat n_{im\sigma} \hat n_{im'\sigma'}\rangle - F^s_\mathrm{DC}.
\end{eqnarray}
Here $\delta_s \equiv d/d \bf{R}_s$ denotes the first-order derivative with respect to the atomic
position $\bf{R}_s$, and $F^s_\mathrm{LDA}$ is the force acting on the atom $s$ calculated within LDA.
Furthermore, $\delta_s \langle {\hat H_\mathrm{LDA}} \rangle$ is evaluated as the thermal average of the
force operator $\delta_s \hat H_{LDA}$, which yields the Hellmann-Feynman contribution
due to the first-order changes of the LDA Wannier Hamiltonian $\hat H_{LDA}$, plus the term due
to the explicit dependence of the local Green function on the atomic positions:
\begin{eqnarray}
\label{eq:kinetic_force}
\delta_s \langle {\hat H_\mathrm{LDA}} \rangle &=& \langle { \delta_s \hat H_\mathrm{LDA}} \rangle \notag \\
&+&
\mathrm{Tr}\sum_{{\bf k},i\omega_n} { \hat H^{\bf k}_\mathrm{LDA} \delta_s \hat G_{\bf k}(i\omega_n) e^{i\omega_n0+}}.
\end{eqnarray}
The first-order derivative of the local Green function is found as
\begin{eqnarray}
\label{eq:gf_derivative}
\delta_s \hat G_{\bf k}(\omega) =
\hat G_{\bf k}(\omega) [ \delta_s \hat H^{\bf k}_\mathrm{LDA}
+ \delta_s \hat \Sigma(\omega) - \delta_s \mu ] \hat G_{\bf k}(\omega).
\end{eqnarray}
Interatomic forces due to the Coulomb interaction, the 4-th term on the right-hand side 
of Eq.~\ref{eq:force_functional}, can be calculated~\cite{Migdal} by using, for example, 
the derivative of the Galitskii-Migdal formula
$ \delta_s E_\mathrm{U} = \frac{1}{2} \mathrm{Tr} \sum_{i\omega_n}
[ \delta_s \hat \Sigma(i\omega_n) \hat G(i\omega_n) +
  \hat \Sigma(i\omega_n) \delta_s \hat G(i\omega_n)] e^{i\omega_n0+}$.
We assume here that the average Coulomb interaction $\bar U$ and Hund's rule coupling
$J$ remain constant when the  atomic positions change. It turns out that the force 
operator $\delta_s \hat H_{LDA}$ and the first-order change of the self-energy 
$\delta_s \hat \Sigma (\omega)$ are the two independent variables in the force 
functional (Eq.~\ref{eq:force_functional}) which have to be evaluated to compute 
interatomic forces \cite{chem_pot}.

To obtain $\delta_s \hat H_{LDA}$, we need to generalize the projection scheme used to evaluate
the LDA Wannier Hamiltonian \cite{AK05,TL08}. The former is based on the projection of
the set of site-centered atomic-like trial-orbitals $|\phi_n \rangle$ on the Bloch functions
$|\psi_{ik} \rangle$ of the chosen bands with band indices $N_a$ to $N_b$. Therefore,
the force operator can be expressed as
\begin{eqnarray}
\label{eq:force_operator}
(\delta_s \hat H^{\bf k}_\mathrm{LDA})_{nm} &=& \sum_{i=N_a}^{N_b} \langle \phi_n|\psi_{i \bf{k}} \rangle \langle \psi_{i \bf{k}}|\phi_m \rangle \notag \\
&&\times (\delta_s V_{i \bf{k}}^\mathrm{KS} + \delta_s V_{i \bf{k}}^\mathrm{Hxc}),
\end{eqnarray}
where $\delta_s V_{i \bf{k}}^\mathrm{KS}$ and $\delta_s V_{i \bf{k}}^\mathrm{Hxc}$
denote the first-order changes in the LDA Kohn-Sham and the Hartree and exchange-correlation
potentials, respectively \cite{basis}. The Kohn-Sham contribution $\delta_s V_{i \bf{k}}^\mathrm{KS}$
can be calculated within the plane-wave pseudopotential approach \cite{RevPWSCF} as
\begin{eqnarray}
\label{eq:analitic_force_operator}
\delta_s V_{i \bf{k}}^\mathrm{KS} &\propto& -i \sum_{{\bf G},{\bf G'}} c^*_{i,{\bf k}+{\bf G}}c_{i,{\bf k}+{\bf G'}}
e^{-i({\bf G}-{\bf G'}) {\bf R}_s} \notag \\
&&\times ({\bf G}-{\bf G'}) V_s^\mathrm{KS}({\bf k}+{\bf G},{\bf k}+{\bf G'}),
\end{eqnarray}
where $V_s^\mathrm{KS}({\bf G},{\bf G'})$ is the Kohn-Sham potential for the atom $s$ (for details
see Ref.~\cite{DC93}). The contribution  $\delta_s V_{i \bf{k}}^\mathrm{Hxc}$ is obtained from
linear-response LDA calculations \cite{DC93}.

To evaluate the change of the self-energy $\delta_s \hat \Sigma(\omega)$ we perform 
a functional derivative of the impurity Green function (here we drop the spin/orbital 
indices and assume summation over repeated indices)
\begin{eqnarray}
\label{eq:variational_approach}
\delta_s \hat G(\tau_1-\tau_2) = - \hat \chi(\tau_1,\tau_2,\tau_3,\tau_4)~
\delta_s \hat {\mathcal{G}}^{-1}(\tau_3,\tau_4)
\end{eqnarray}
with
\begin{eqnarray}
\label{eq:susceptibility}
\hat \chi(\tau_1,\tau_2,\tau_3,\tau_4) = \langle \mathcal{T}_{\tau} \hat c(\tau_1) \hat c^{\dagger}(\tau_2) \hat c^{\dagger}(\tau_3) \hat c(\tau_4)\rangle - \notag \\
\langle \mathcal{T}_{\tau}  \hat c(\tau_1) \hat c^{\dagger}(\tau_2) \rangle
\langle \mathcal{T}_{\tau}  \hat c^{\dagger}(\tau_3) \hat c(\tau_4) \rangle,
\end{eqnarray}
and use the first-order derivative of the local Green function (Eq.~\ref{eq:gf_derivative}).
We solve Eqs.~\ref{eq:gf_derivative} and \ref{eq:variational_approach} self-consistently by 
employing $\delta_s \hat {\mathcal G}^{-1} = \delta_s \hat G^{-1} + \delta_s \hat \Sigma$ and
the two-particle correlation function, i.e., the generalized susceptibility,
$\chi(\tau_1,\tau_2,\tau_3,\tau_4)$ calculated within DMFT.
The proposed method \cite{selfconsistency} is implemented using the Hirsch-Fye
quantum Monte Carlo (QMC) method \cite{HF86}.

\begin{figure}[tbp!]
\centerline{\includegraphics{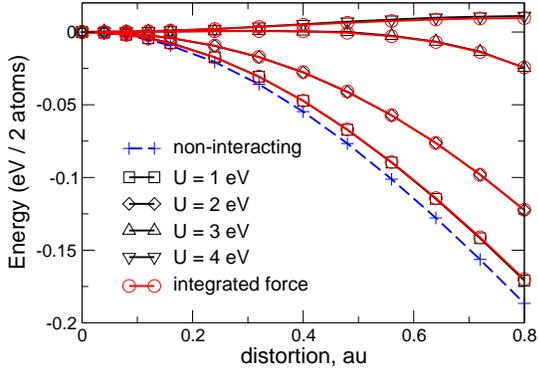}}
\caption{(color online)
Comparison of the total energies of paramagnetic hydrogen computed by LDA+DMFT with the
results obtained by numerical integration of the corresponding force with respect to an 
atomic displacement. The calculations are performed using different values of Coulomb
interaction $U$.
} \label{fig:H_total_force}
\end{figure}

We now perform several test calculations to illustrate how the proposed method 
works in practice. To demonstrate its accuracy, we compare our results for
the total energy calculated as a function of atomic displacement with those
obtained by numerical integration of the corresponding forces. As the first 
test, we consider the simplest correlated electron problem, elemental hydrogen 
(H), with a cubic structure and lattice constant $a=8$ atomic units (a.u.).
The nonmagnetic LDA calculations for cubic hydrogen yield a metallic solution with
a half-filled H $s$ band of 3 eV width located at the Fermi level.
To evaluate the force, we consider a supercell with two hydrogen atoms,
in which one of the atoms is displaced by a distance $\delta$ with respect to its 
crystallographic position. In Fig.~\ref{fig:H_total_force} we present our results 
for the total energy obtained by LDA as a function of $\delta$. The nonmagnetic 
LDA calculations find the cubic lattice of hydrogen to be unstable since the 
total energy decreases with $\delta$.

\begin{figure}[tbp!]
\centerline{\includegraphics{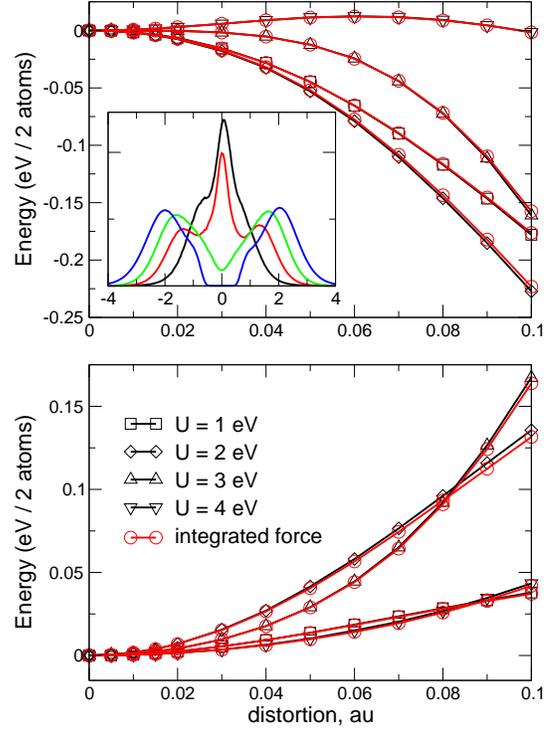}}
\caption{(color online)
Kinetic (top) and interaction (bottom) contributions to the total energy of paramagnetic
hydrogen as calculated by LDA+DMFT in comparison with the results obtained by numerical 
integration of the corresponding force. Inset: Evolution of the resulting spectral 
function as a function of $U$.
} \label{fig:H_partial_forces}
\end{figure}

Now we take into account the electronic correlations by calculating the properties
of paramagnetic hydrogen using the LDA+DMFT method. For the partially filled H $s$
orbitals  a basis of atomic-centered symmetry constrained Wannier functions is constructed.
By calculating at different values of the local Coulomb interaction $U$ we can explore the 
structural properties of correlated materials near a Mott metal-insulator phase 
transition --- a challenging problem in solid state research.
The calculations are performed for the $U$ values in the range of 1-4 eV
at a temperature $T = 0.1$ eV. Our results for the spectral function of
paramagnetic hydrogen with $\delta=0$ are shown in the inset of
Fig.~\ref{fig:H_partial_forces}.

In Fig.~\ref{fig:H_partial_forces} we present our results for the total
energies calculated by LDA+DMFT for paramagnetic hydrogen as a function 
of the displacement $\delta$. By changing the $U$ values, we are able to 
check the accuracy of our method in calculating the kinetic and interaction 
contributions, respectively, to the total force. By integrating the 
corresponding force  with respect to $\delta$, we find an overall good 
quantitative agreement (within 1-2 meV) between the force and the total 
energy calculations. Even for large displacements $\delta$ (up to $\sim$ 
10 \% of the lattice constant $a$) our force calculations show an excellent 
accuracy of $\leq 1$ mRy/a.u. in the whole range of the $U$ values.
In Fig.~\ref{fig:H_total_force} we provide a comparison of the total energy 
with the results of the numerical integration of the corresponding total force. 
Most interestingly, by increasing $U$, the cubic lattice (more precisely, the 
investigated displacive mode) becomes (meta-) stable for $U \geq 4$ eV.
These results clearly demonstrate the crucial importance of electronic 
correlations for the lattice stability of correlated materials.

\begin{figure}[tbp!]
\centerline{\includegraphics{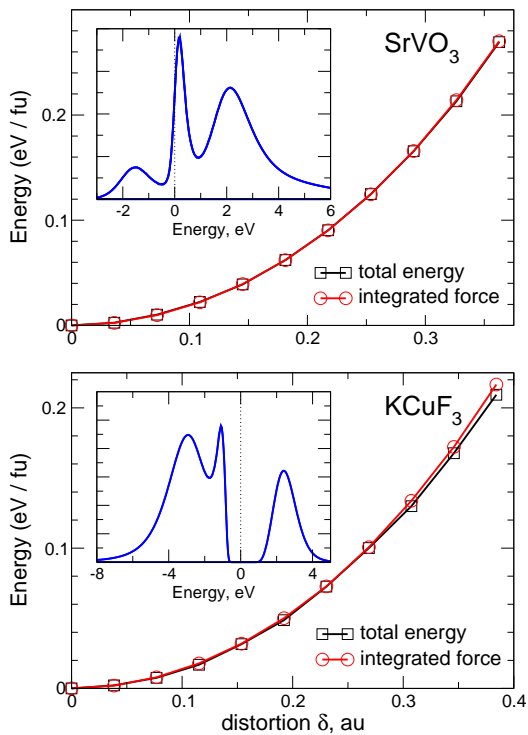}}
\caption{(color online)
Comparison of the total energies of paramagnetic SrVO$_3$ (top) and KCuF$_3$ (bottom) 
computed by LDA+DMFT with the results obtained by numerical integration of the 
corresponding forces with respect to the oxygen and fluorine displacement, respectively.
} \label{fig:materials}
\end{figure}

Next we apply our linear-response method to investigate a realistic
correlated electron system, SrVO$_3$. This material has a cubic perovskite
structure and a V $3d^1$ electronic configuration. According to previous
electronic-structure studies SrVO$_3$ is a strongly correlated metal,
with a well established three-peak structure in the spectral function
\cite{SrVO3}. It exhibits pronounced lower and upper Hubbard bands, which
cannot be explained by conventional LDA. SrVO$_3$ is an ideal test material 
to benchmark our computation of forces and, thereby, the prediction of the 
atomic positions.
In our calculations we use the experimental cubic structure with lattice
constant $a = 3.838$ \AA\ and take the Coulomb interaction $\bar U=3.55$ eV
and exchange coupling $J=1.0$ eV from previous constrained LDA calculations 
\cite{SrVO3}. For the partially filled V $t_{2g}$ orbitals a basis of atomic-centered
symmetry constrained $t_{2g}$ Wannier functions is constructed. In Fig.~\ref{fig:materials}
(upper panel) we present our results for the spectral function of paramagnetic SrVO$_3$ 
obtained by LDA+DMFT for $T = 0.125$ eV. Overall, our results qualitatively agree
with previous calculations.

In order to model the internal lattice distortions, we displace the apical 
oxygen atom O $(0 0 \frac{1}{2})$  by $\delta$ in the $z$-direction, thereby
introducing a polar mode. In Fig.~\ref{fig:materials} (upper panel)
we present our results for the total energy obtained by LDA+DMFT
as a function of $\delta$. These calculations predict the polar mode to be
unstable, implying the internal stability of the cubic perovskite structure 
of SrVO$_3$. We evaluate the LDA+DMFT forces acting on that particular oxygen 
atom as a function of $\delta$. By integrating the force, the accuracy of
our method in predicting the correct atomic positions can be checked.
We find excellent agreement (within 1 meV/fu) between the force and the
total energy calculations. Our results clearly show that our method is able
to treat atomic displacements for a realistic correlated electron metal
such as SrVO$_3$.

Finally we turn to KCuF$_3$, a prototypical Mott-Hubbard insulator with
a single hole (Cu $3d^9$ electronic configuration) in the Cu $e_g$ states.
For simplicity, a hypothetical cubic perovskite lattice is assumed, with
$a = 4.066$ \AA, i.e. the Jahn-Teller distortion and tetragonal compression 
of the unit cell are neglected. Constructing a basis of atomic-centered 
symmetry constrained $e_g$ Wannier functions for the partially filled Cu 
$e_g$ orbitals, we calculate the electronic structure of paramagnetic KCuF$_3$
within the LDA+DMFT approach for $T = 0.125$ eV, using the interaction parameters 
$\bar U=5.2$ eV and $J=0.9$ eV from previous calculations \cite{LB08}. 
The obtained spectral function is shown in Fig.~\ref{fig:materials} (bottom 
panel). Similar to SrVO$_3$, we introduce a polar mode by shifting an apical 
fluorine atom F $(0 0 \frac{1}{2})$ in the $z$-direction. By integrating
the resulting force, our results  can be compared with the total energy calculations.
We find an overall good quantitative agreement, implying internal consistency
and numerical stability of the proposed linear-response approach. Our results 
for both materials, the correlated metal SrVO$_3$ and the correlated Mott-Hubbard 
insulator KCuF$_3$, demonstrate that the linear-response method presented here provides
a robust computational tool for the study atomic displacements caused by electronic 
correlations. In particular, it allows one to determine the structural
phase stability of both metallic and insulating correlated materials in
their paramagnetic and magnetically ordered state.


In conclusion, by implementing LDA+DMFT with the linear-response formalism regarding
atomic displacements, we constructed a robust computational scheme for the investigation
of the electronic structure and lattice properties of correlated electron materials.
The approach allows one to calculate forces and thereby explore lattice transformations
induced by electronic correlations. In particular, it is now possible to study lattice
instabilities observed at correlation induced metal-insulator transitions. Furthermore,
lattice dynamical properties of correlated electron materials can be calculated by 
implementing the approach with, for example, the so-called small displacements method
\cite{DA09}.

\begin{acknowledgments}
We thank N. Binggeli, Dm. Korotin, and J. Kune\v{s} for valuable discussions. Support by the
Deutsche Forschungsgemeinschaft through TRR 80 (I.L.) and FOR 1346 (V.I.A., D.V.), as well
as by RFFI 13-02-00050 is gratefully acknowledged.
\end{acknowledgments}


\begin{thebibliography}{99}
\bibitem{Migdal}
The derivative $\delta_s E_\mathrm{U}$ can be also evaluated using a variation of the double
occupancy matrix $\delta_s \langle \hat n_{im\sigma} \hat n_{im'\sigma'}\rangle$. However, 
this is computationally more demanding, since it requires the  calculation of a three-particle 
correlation function.
\bibitem{basis}
We assume here that the Wannier basis is complete, i.e., there is no contribution
caused by the change of the projectors $\delta_s \langle \phi_n | \psi_{i \bf{k}}\rangle $.
In general,
$\delta_s \langle \phi_n | \psi_{i \bf{k}}\rangle $ can be evaluated using the orthonormality
conditions
$\sum_{i=N_1}^{N_2}\langle \phi_n | \psi_{i \bf{k}}\rangle \langle \psi_{i \bf{k}} | \phi_m\rangle = \delta_{nm}$.
\bibitem{selfconsistency}
Here we perform DMFT calculations for a fixed LDA Hamiltonian $\hat H_{LDA}$,
thereby neglecting  full charge self-consistency.
\end{thebibliography}
\end{document}